\documentclass[aps,prb,twocolumn,showpacs,superscriptaddress]{revtex4}

\usepackage[cp866]{inputenc}
\usepackage[russian]{babel}
\usepackage{bm}
\usepackage{graphicx}

\newcommand{\beq}{\begin{equation}}
\newcommand{\eeq}{\end{equation}}
\begin{document}

\title{\boldmath
Conventional BCS, Unconventional BCS, and Non-BCS Hidden Dineutron  Phases in
Neutron Matter}
\author{V.~A.~Khodel}
\affiliation{National Research Centre Kurchatov
Institute, Moscow, 123182, Russia}
\affiliation{McDonnell Center for the Space Sciences \&
Department of Physics, Washington University,
St.~Louis, MO 63130, USA}
\author{J.~W.~Clark}
\affiliation{McDonnell Center for the Space Sciences \&
Department of Physics, Washington University,
St.~Louis, MO 63130, USA}
\affiliation{Centro de Ci\^encias Matem\'aticas\\ University of
Madeira, Funchal, Madeira, Portugal}
\author{V.~R. Shaginyan}
\affiliation{Petersburg Nuclear Physics Institute, NRC Kurchatov
Institute, Gatchina, 188300, Russia}
\author{M.~V.~Zverev}
\affiliation{National Research Centre Kurchatov
Institute, Moscow, 123182, Russia}
\affiliation{Moscow Institute of Physics and Technology, Dolgoprudny, Moscow District 141700, Russia}

\date{\today}
\begin{abstract}
The nature of pairing correlations in neutron matter is re-examined.
Working within the conventional approximation in which the $nn$
pairing interaction is provided by a realistic bare $nn$ potential
fitted to scattering data, it is demonstrated that the standard BCS
theory fails in regions of neutron number density where the pairing
constant $\lambda$, depending crucially on density, has a non-BCS
negative sign. We are led to propose a non-BCS scenario for pairing
phenomena in neutron matter that involves the formation of a
hidden dineutron state. In low-density neutron matter where the
pairing constant has the standard BCS sign, two phases organized by
pairing correlations are possible and compete energetically: a conventional BCS phase and a dineutron phase. In dense neutron matter, where $\lambda$ changes sign, only the dineutron phase survives and exists until the critical density for termination of
pairing correlations is reached at approximately twice the neutron
density in heavy atomic nuclei.

\end{abstract}

\maketitle

\section*{PREAMBLE}
This contribution is dedicated with deep respect and admiration to
Spartak Timofeevich Belyaev on the occasion of his 90th birthday.
Three generations of physicists across the globe have taken inspiration
from his prodigious achievements in theoretical nuclear physics and
quantum many-body theory, as well as his wise and visionary leadership
in the development and sustenance of world-renowned scientific institutions.
Recognized by awards of the 2004 Feenberg Memorial Medal and the 2012
Pomeranchuk Prize as one of the founding fathers of modern many-body
theory based on field-theoretic methods, he introduced the concept of
anomalous propagators\cite{belb} that runs through all of current theoretical
physics and is central to a microscopic understanding of pairing phenomena
in nuclear and condensed matter systems. The impact of the profound advances
he made in the theory of nuclear superfluidity during his 1957-58 ``wonder
year'' at the Bohr Institute and later in Novosibirsk, changed the course of
nuclear theory, giving rise to the ``standard nuclear paradigm'' in which
BCS pairing correlations assume pivotal roles.\cite{belc,belz}
S.~T.\ has taught us\cite{bel50} that ``There is still a vast field
of unsolved problems stimulating the progress of theoretical nuclear
physics.'' Our contribution to this celebratory issue is offered very
much in the same spirit, as we seek to establish that the
study of pairing correlations remains a source of surprising
and intriguing revelations about the microworld.

\section{Introduction}
Shortly after Bardeen, Cooper, and Schrieffer (BCS) introduced a
theory of superconductivity in 1957, A.~B.~Migdal raised the possibility
that the matter inside neutron stars may be superfluid.  Since that time,
hundreds of papers have been published to elucidate the properties of
neutron matter and other nuclear systems implied by nucleonic pairing,
within the framework of BCS theory.\cite{BZbook}  In the generic
zero-temperature Lifshitz phase diagram of a homogeneous 3D Fermi
system subject to pairing correlations, the conventional BCS phase
lies in the weak-coupling domain of small {\it positive} pairing
constant $\lambda$.  Specifically, this dimensionless coupling
parameter is defined by $\lambda=-V_FN(0)$, where $V_F=V(p_F,p_F)$
is the diagonal matrix element of the pairing interaction and
$N(0)=p_FM^*/\pi^2$ is the density of single-particle states,
both evaluated at the Fermi surface. (The Fermi momentum
is given by $p_F=(3\pi^2\rho)^{1/3}$ in terms of the particle
density $\rho$, while $M^*$ stands for the effective mass.)
The occurrence of the BCS phase in this domain is attributed to
the enhancement of pairing correlations stemming from the
logarithmic divergence of the propagator of a pair of
opposite-spin quasiparticles as their total momentum $\bf P$
approaches zero.  This enhancement leads to the formation of a
condensate of Cooper pairs with ${\bf P} = 0$, which entails
violation of global U(1) phase rotation symmetry,
and is responsible for the superfluidity of the BCS phase.
A crucial feature of this phenomenon is the presence of
a gap $\Delta(p)$ in the spectrum $E(p)$ of single-particle
excitations.  In the relevant region of the Lifshitz phase
diagram, the value of the BCS gap $\Delta_0\equiv \Delta(p=p_F,T=0)$
and critical temperature $T_c$, above which the BCS gap closes
and BCS superfluidity is terminated, turn out to be
exponentially small:
\beq
\Delta_0=\Omega_D e^{-2/\lambda},
\quad  T_c = 0.57 \Delta_0 ,
\label{bcsres}
\eeq
where $\Omega_D$ is the BCS cutoff factor.

BCS theory reigned for several decades as the most successful theory in
condensed-matter physics, both fundamentally and quantitatively.
However, its limitations became apparent after the discovery of a
family of high-temperature superconductors in the late 1980's.  Failure
of the theory was conclusively established with the revelation of the
so-called {\it pseudogap phase} in experimental studies of putatively normal
phases of high-$T_c$ superconductors by means of angular-resolved-photoemission
spectroscopy (ARPES).  In such a phase, there still exists a gap in the
single-particle spectrum, even though the superconductivity is already
terminated.\cite{timusk,shen}  BCS theory, a bedrock of our understanding
of the phenomena of superfluidity and superconditivity in which
termination of these phenomena and closure of the energy gap are
inseparable, is manifestly inappropriate when we attempt to describe
the pseudogap phase.

A plethora of scenarios have been offered in explanation of such challenging
behavior of high-$T_c$ superconductors.  Their discussion is well beyond
the scope of the present article, in which we choose to highlight a scenario
associated with the original model of in-medium pairing correlations
explored by Shafroth, Butler, and Blatt\cite{shafroth,sbb,blatt} in the years
leading up to the breakthrough made by BCS.  This scenario envisions
the formation of bound pairs {\it in real three-dimensional
space}.\cite{blatt,alexr,nozs,roepke,lombardo,alexm,alm,alexb}
Such a process becomes feasible
in the strong-coupling limit when the pair radius turns out to be smaller
than the mean interparticle distance, while the pair binding energy
${\cal E}$, playing the role of a gap in the spectrum of single-particle
excitations, exceeds the Fermi energy $\epsilon^0_F=p^2_F/2M$.

It follows that the pairing phase thus envisioned should involve the
phenomenon of Bose-Einstein (BE) condensation. The most fully developed
treatment of this phenomenon in solid-state physics, known as the theory
of bipolaronic superconductivity, is the pioneering work of the
late A.\ S.\ Alexandrov and his coauthors.\cite{alexr,alexm,korn}
To honor his contribution, we call this phase of matter the
Shafroth-Butler-Blatt-Alexandrov (SBBA) phase. The scenario of
bipolaronic superconductivity is based on the polaron concept as
set forth by Landau in 1933.\cite{lanp,pekar0}  Conventional polarons,
having spin $1/2$, result from interactions between electrons and
optical phonons, their mass $M_p$ appearing to be much larger the
electron mass $M$.\cite{pekar,korn}  It is the mass $M_p$ that enters the
criterion for creation of a bound state of two polarons, the so-called
bipolaron, and this criterion is met even if the attraction between
polarons is moderate.  In the description of superconductivity as a
BE condensation of bound electron pairs, an idea already advanced
by London in 1938, the interplay between bound pairs and the
continuum of two-particle states is treated theoretically within
the concept of quasichemical equilibrium, in analogy to thermodynamics
of ordinary chemical reactions as presented in textbooks.

As $T$ increases, the density of the superfluid Bose-Einstein condensate
of real-space pairs declines and eventually vanishes, terminating
superfluidity.  The critical temperature $T_c^{BE}$ for destruction of
bipolaronic superconductivity is not exponentially small as
in Eq.~(\ref{bcsres}), instead showing qualitative agreement with
the behavior observed in high-$T_c$ superconductors.  Since
SBBA theory attributes the property of superfluidity to the
bosonic system of bound pairs, there should be no jump of the specific
heat $C(T)$ at $T=T_c^{BE}$, in contrast to this distinctive
signature of BCS pairing at the associated critical temperature.
Furthermore, it is easily verified that in SBBA theory the density
of unbound fermions is proportional to $e^{-{\cal E}(T)/T}$.  Hence
their contribution can be safely ignored when ${\cal E}(T_c) \gg T_c$,
and the ARPES data then give evidence for the persistence of the
gap $\Delta(T)\propto {\cal E}(T)$ in the spectrum of single-particle
excitations above $T_c$.

The crisis faced by the BCS description in dealing with strongly correlated
electron systems of solids, happening after 50 years of serenity, calls
equally for a reassessment of the theory of {\it nuclear} pairing
correlations, since nuclear systems, including atomic nuclei and
neutron matter, are also composed of strongly correlated fermions.

In neutron matter, there exists a potential nuclear analog of the
bipolaron, the {\it in-medium dineutron}.  Like the bipolaron,
the dineutron is non-existent in vacuum.  However, analogously to
the bipolaron situation, the presence of the background medium might
promote the formation of bound dineutron pairs.  Highly relevant to
this possibility is a distinctive feature of neutron-neutron scattering,
namely a narrow resonance lying at the tiny energy of 0.067 MeV, which
implies that neutrons attract each other much more effectively by
intrinsic nuclear forces than do two electrons by means of phonon exchange.

The dineutron state exhibits itself as a pole in the Cooper channel
of the in-medium $nn$ scattering amplitude, quite unrelated to the
Fermi surface and existing until the density $\rho$ reaches a critical
value $\rho_t\simeq 2\rho_0\simeq 0.16 ~ {\rm fm}^{-3}$,
where $\rho_0$ denotes the neutron density in heavy nuclei.
Alas, the standard BCS approach is legitimate only at low densities
$\rho\leq 0.4\rho_0$.  At larger densities, the pairing constant
$\lambda$, being critically dependent on $\rho$, changes sign
(see Fig.~1), and application of the BCS theory becomes questionable.

Notwithstanding the subtle complexity and inherent richness of fermionic
pairing, studies of the implications of pairing correlations over the
last half century have been uniquely pursued within standard BCS theory.
This article represents a first step toward understanding the details of
the interplay between BCS and dineutron pairing correlations in
neutron matter.

\section{The state of the art}

In the decades following the landmark BCS paper, theorists achieved many
successes in quantitative treatment of the pairing interaction between
electrons in solids, as derived from electron-phonon exchange between
electrons with energies near the Fermi surface.
By contrast, the pairing problem in more strongly correlated Fermi systems
such as neutron matter and liquid $^3$He continues to present serious
challenges to quantitative, {\it ab initio} microscopic description,
in spite of numerous efforts in this direction.\cite{jwcbcs,baldo}

Here we will deal with fundamental aspects of this more difficult
class of systems that have previously escaped recognition, namely
the possibility of a non-BCS pair-organized phase of neutron matter
 based on dineutron formation.  In doing so, we will focus on
qualitative rather than quantitative issues.  Within this limited
objective, it is reasonable to adopt the conventional approximation
in which the block of Feynman diagrams irreducible in the Cooper
channel is replaced by a vacuum $nn$ interaction potential $V$
of phenomenologically motivated form, fitted to two-nucleon scattering
data.  The Reid soft core (RSC) potential\cite{reid}
is one such interaction that remained popular among nuclear theorists
over an extended period.  Highly refined potential models of the same
type are in current use; prominent among these are the Argonne
$V_{14}$ and $V_{18}$ potentials.\cite{wir}

\begin{figure}
\scalebox{0.50}
{\includegraphics{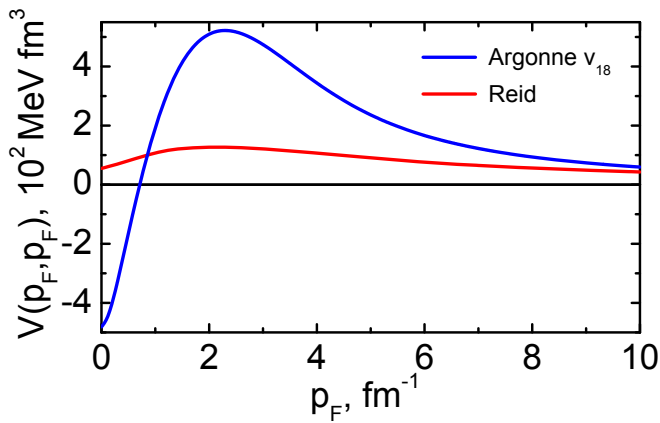}}
\caption{Diagonal two-body pairing matrix element at the Fermi surface,
$V(p_F,p_F)$, versus Fermi momentum $p_F$, plotted for
Reid soft-core (RSC) and Argonne $V_{18}$ potential models.}
\label{figure_vF}
\end{figure}

As seen from Fig.~\ref{figure_vF}, the RSC interaction has the
interesting property that the pairing constant $\lambda$, which has
a positive sign in BCS theory, remains {\it negative} for all values
of $p_F$, due the strong inner repulsion present in this potential
model.  For the currently popular $nn$ potential models mentioned
above, notably $V_{14}$ and $V_{18}$, $\lambda$ is seen to have
the conventional positive sign at small $p_F$ values.  However,
this coupling parameter again shows a strong negative excursion as
the density increases beyond   $p_F\simeq 0.8$ fm$^{-1}$.
Of course, for a
solution of the BCS gap equation to exist, the pairing
interaction $V$ must take negative values for some range of particle
separations $r$ in {\it coordinate space}, corresponding to attraction
between quasiparticles, and the RSC potential and other more realistic
bare $nn$ interaction models certainly do meet this requirement.  Thus
it is no surprise that solutions of the gap equation for the RSC
potential do exist, yielding a substantial maximum of the gap
value $\Delta_0$ close to 3 MeV at $p_F \approx 0.85$ fm$^{-1}$
(e.g., see Ref.~\onlinecite{kkc}).

Consequent to this behavior, solutions of the BCS gap equation obtained
for realistic $nn$ potentials exhibit a striking feature relative to
the conventional BCS scenario.  According to Eq.~(\ref{bcsres}), the gap
value $\Delta_0$ should increase rapidly and monotonically with
increasing particle density, since the density of states $N(0)$ entering
this formula is proportional to $p_F$.  In the case of neutron matter
described by the class of $nn$ potentials studied, $\Delta_0$ increases
with $p_F$ up to a maximum around $0.85$ fm$^{-1}$, then falls off and
eventually closes at the critical density
$\rho_t=p^3_{Ft}/3\pi^2$
corresponding to a Fermi momentum $p_{Ft} \simeq 1.74~{\rm fm}^{-1}$.
(Numerical values are cited for the RCS potential; very similar
results are obtained for the more modern potential models.)
This feature can be ascribed to the occurrence of a bifurcation point
in the BCS gap equation when the pairing interaction is constructed from
a realistic $nn$ potential.

\section{Two types of pairing instability of the normal state in
neutron matter}

We shall use the term ``conventional BCS solutions'' to designate
solutions existing in the case $ \lambda>0$ that behave in accordance
with Eq.~(\ref{bcsres}) when $\lambda$ tends to 0.  The pairing solutions
obtained for the RSC potential in Ref.~\onlinecite{kkc}, and in numerous
independent calculations for the Argonne potentials, must then be
identified as {\it unconventional}
solutions of the BCS gap equation, since the associated pairing constants
$\lambda$ have the ``wrong'' (i.e., negative) sign over an extensive density
range.  In this connection, it is significant that in the density regime
relevant to our discussion, the BCS gap function $\Delta(p)$ found for such
potentials practically coincides with that of the dineutron solution of
the Schr\"odinger equation in momentum space, provided that the neutron
mass is only slightly enhanced (by $\simeq 5\%$) so as to admit a bound
dineutron pair.  This fact suggests the presence of a hidden dineutron
state that is responsible for elements of the unorthodox behavior
of the gap amplitude $\Delta_0(\rho)$ in superfluid nuclear matter.

To clarify the situation it is expedient to trace the location of the
Cooper singlet-channel pole of the zero-temperature scattering
amplitude $\Gamma({\bf P}=0,\omega)$ in the normal state of the system.
This can be done based on the Bethe-Salpeter (BS) equation for the
corresponding vertex part
${\cal T}_{\alpha\beta}({\bf p},\omega)\equiv {\cal T}({\bf p},\omega)(\tau_2)_{\alpha\beta}$, where
$\alpha,\beta$ are spin indices, ${\bf p}$
is the momentum of the incoming quasiparticle (with its target having
momentum $-{\bf p}$), $\omega$ is the total two-particle energy measured
from $2\mu$, and $\mu$ is the chemical potential.
In our treatment, the required equation\cite{trio,lo} reads
\beq
{\cal T}(p,\omega)=-\int V_0( p,p_1)L(p_1,\omega)
{\cal T}(p_1,\omega)d\upsilon_1
\label{gc}
\eeq
in terms of the zeroth harmonic $ V_0( p, p_1)$ of the
interaction potential $V$ (which replaces a block of diagrams
irreducible in the particle-particle channel), the particle-particle
propagator $L(p,\omega)$ of the normal ground state, given by
\beq
L(p,\omega)= -(1-2n(p))/(\omega-2\epsilon(p)-i\delta {\rm sgn} (p-p_F)),
\label{prbcs}
\eeq
with $\epsilon(p)$ the single-particle spectrum (chosen to coincide with the
bare spectrum $\epsilon(p)=p^2/2M-\mu$ and $d\upsilon$ the volume element
of 3D momentum space.

The central problem encountered in the analysis of Eq.~(\ref{gc}),
as well as more complicated nonlinear integral equations of the
theory of pairing correlations, lies in the presence of the Cooper
singularity in the two-dimensional kernel.  A method developed and
implemented for the BCS problem in Ref.~\onlinecite{kkc} (and
discussed in more detail in Ref.~\onlinecite{khod1997}) serves
to isolate the singularity and thereby overcome this problem in two
relatively easy stages.  First, the pairing interaction is decomposed
into a separable part and a remainder, so as to derive a linear
integral equation for the momentum dependence of the solution in
which the presence of this singularity is immaterial.  Once this
linear equation is solved, we are left with a nonlinear equation
for a gap amplitude or other related quantity, whose analysis and numerical
solution are far simpler (and more accurate) than in direct treatment
of the original nonlinear equation.  As will be seen, this approach also
proves advantageous in the forthcoming disclosure of a hidden dineutron
phase in neutron matter subject to pairing correlations.

The zeroth harmonic $V_0$ entering Eq.~(\ref{gc}) is decomposed
as follows
\beq
V_0(p_1,p_2)= V_F\phi(p_1)\phi(p_2)+ {\cal R}(p_1,p_2),
\label{decg}
\eeq
where $\phi(p)= V_0(p,p_F)/ V_F$ with $ V_F= V_0(p_F,p_F)$ and hence
$\phi(p_F)=1$. This decomposition is designed to yield the property
\beq
{\cal R}(p,p_F)={\cal R}(p_F,p)=0 .
\label{regr}
\eeq
Upon insertion of Eq.~(\ref{decg}) into Eq.~(\ref{gc}) followed by
some algebra, we are led to a set of two coupled equations.  The first
of these,
\beq
D(p,\omega)=\phi(p)-\int {\cal R}(p,p_1)
L(p_1,\omega)D(p_1,\omega)d\upsilon_1,
\label{gam}
\eeq
is an equation for the shape factor $D(p)\equiv {\cal T}(p)/{\cal T}(p_F)$,
which is almost unaffected by the Cooper singularity because the
remainder ${\cal R}(p,p_1)$ {\it vanishes identically} when either
argument is on the Fermi surface.  The second equation,
\beq
-1/V_F=\int \phi(p)L(p,\omega)D(p,\omega)d\upsilon,
\label{om}
\eeq
determines the location of the pole itself.

In the standard BCS situation with the Debye frequency
$\Omega_D\ll \epsilon^0_F$, the remainder ${\cal R}$ is suppressed.
Analytically continuing Eq.~(\ref{om}) into the complex
$\omega$ plane, we can therefore employ the first approximation
$D^{(1)}(p)=1$ to find
\beq
  {1\over \lambda}=0.5\left(\ln {\Omega_D\over \omega}+i{\pi\over 2}\right),
\label{bcd}
\eeq
where $\lambda=-N(0)V_F$ as before.
This equation has the solution $\omega=i\Omega$, where the real
number $\Omega$ is found from the BCS equation
\beq
{1\over\lambda}=0.5\,\ln {\Omega_D\over\Omega}  ,
\eeq
implying that one is dealing with the standard Cooper instability,
which is eliminated through formation of the Cooper condensate.
In the RSC case with $ \lambda<0$, the approximation $D^{(1)}(p)=1$
fails: the dispersion equation (\ref{bcd}) has no solutions at all,
at variance with numerical results obtained from an iterative procedure
or different methods of solving the standard BCS gap equation.

Pursuant to the point, let us address the case of small $\omega\to 0$
and recast Eq.~(\ref{om}) in the form
\beq
-1/ V_F=I_{11}(\omega)
+\int \phi(p)L(p,\omega)\eta(p,\omega)d\upsilon  ,
\label{far1}
\eeq
wherein
\beq
I_{11}(\omega)=\int \phi^2(p)L(p,\omega)d\upsilon=
  0.5\,N(0)\left(\ln{\epsilon_c\over\omega}+i{\pi\over 2} \right) ,
\label{i11}
\eeq
with a cut-off energy $\epsilon_c$,
while the function $\eta(p,\omega)=D(p,\omega)-\phi(p)$, determined
at arbitrary $\omega$ and $\rho$, obeys the equation
\beq
\eta(p,\omega)=-\int {\cal R}( p,p_1;\rho)
L(p_1,\omega)(
  \phi(p_1)
+\eta(p_1,\omega))d\upsilon_1 .
\label{far2}
\eeq

Since the neighborhood of the Fermi surface contributes divergently
only to the first integral on the right side of Eq.~(\ref{far1}),
it would seem that nontrivial solutions with small $\Omega$ simply
do not exist when $ V_F>0$.  However, this is not the case.
Such solutions do in fact emerge in the vicinity of a critical
density $\rho_{cr}$ at which the second term on the right side
of Eq.~(\ref{far1}) {\it diverges as well} due to the divergence
of the function $\eta(p,0)$ at the bifurcation point.  Thus,
the singular terms conspire to cancel each other.\cite{kkc,khod1997}
It is known from the theory of integral equations that the solution of
an inhomogeneous linear integral equation such as (\ref{far2}) does
indeed diverge at a critical density $\rho_0$ where the lowest
eigenvalue $\sigma_0$ of the kernel ${\cal R}( p,p_1,\rho)L(p,0)$,
determined from the equation
\beq
\zeta_0(p)=-\sigma_0\int {\cal R}( p,p_1,\rho)L(p_1,0)\zeta_0(p_1)
d\upsilon_1 ,
\label{ztoo}
\eeq
{\it is equal to unity}.

The structure of the diverging component of $\eta(p,\omega)$ is readily
accessible by standard operations.  For consider that the function
$\eta(p,\omega)$ may be expanded in a basis formed by the
eigenfunctions $\zeta_n(p)$ of the above kernel.  Extracting the
main term proportional to $\zeta_0(p)$ explicitly, we may write
\beq
\eta(p,\omega)=\eta_0(\omega)\zeta_0(p)+\vartheta(p)
\label{exo}
\eeq
where the remainder $\vartheta(p)$ vanishes at the Fermi surface
like $\zeta_0(p)$ and $\eta(p)$.  Inserting this formula into Eq.~(\ref{far2})
and gathering all terms explicitly containing the factor $\eta_0(\omega)$
on the left side of the equation, we may arrive at
\beq
\eta_0(\omega)\left (\zeta_0(p)+
\int {\cal R}( p,p_1) L(p_1,\omega)\zeta_0(p_1)d\upsilon_1\right)=Y(p,\omega) ,
\label{etaom}
\eeq
where
\beq
Y(p,\omega)=-\vartheta(p)-\int {\cal R}( p,p_1) L(p_1,\omega)\left(\phi(p_1)
+\theta(p_1)\right)d\upsilon_1 .
\eeq
With the aid of Eq.~(\ref{ztoo}), the left side of Eq.~(\ref{etaom}) is recast in
the form
\beq
\eta_0(\omega)\left ({\kappa\over \sigma_0}\zeta_0(p)+ \int {\cal R}( p,p_1)\delta L(p_1,\omega)\zeta_0(p_1)d\upsilon_1\right)
=Y(p,\omega) ,
\label{echo}
\eeq
where $\delta L(p,\omega)=L(p,\omega)-L(p,0)$.  Here we have also introduced the effective
stiffness coefficient
\beq
\kappa=\sigma_0-1 ,
\label{kap}
\eeq
which is central to the problem under discussion.

At the next step, we multiply both sides of Eq.~(\ref{echo}) by the product
$\zeta_0(p)L(p,0)$ and integrate over the momentum $p$.  Eliminating the
operator ${\cal R}$ in the same way as before, we obtain
\beq
\eta_0(\omega)\left(\kappa+B(\omega)\right)=I_{10}(\omega)/I_{00} ,
\eeq
where the factor $B$ is given by
\beq
B(\omega)=-(I_{00})^{-1}\int \zeta_0(p) \delta L(p,\omega) \zeta_0(p) d\upsilon,
\label{bo}
\eeq
while
\begin{eqnarray}
I_{00}&=&\int \zeta_0(p)   L(p,0)
\zeta_0(p)d\upsilon ,\nonumber \\
I_{10}(\omega)&=&\sigma_0\int \zeta_0(p)
  L(p,0)
 Y(p,\omega)d\upsilon  .
\label{i00}
\end{eqnarray}
Upon substituting the explicit form of the function $Y(p,\omega)$ into the
last of these integrals, it is found that the terms in the remainder
$\vartheta$ practically cancel each other.   We are left with
\beq
I_{10}(\omega)\simeq I_{10}\equiv\int \zeta_0(p) L(p,0)\phi(p)d\upsilon
\eeq
and therefore arrive at
\beq
\eta_0(\omega)={I_{10}/I_{00} \over  \kappa+B(\omega)} .
\label{etome}
\eeq
Since $B(\omega=0)$ vanishes, we may then infer that the coefficient
$\eta_0(\omega=0)$ given by Eq.~(\ref{etome}), and hence the function
$\eta(p,0)$, do in fact diverge at the critical density $\rho_t$
where $\kappa(\rho)$ vanishes.

At the final step, Eqs.~(\ref{exo}) and (\ref{etome}) are inserted into
Eq.~(\ref{far1}).  After deleting insignificant contributions
from the regular term $\vartheta(p)$, we arrive at the
required dispersion equation, whose analytical continuation
to the complex $\omega$ plane has the form
\beq
0.5\left(\ln{\epsilon_c\over \omega}+i\pi/2\right)={1\over \lambda}
-{\nu^2\over \kappa+B(\omega)},
\label{deq}
\eeq
where we have employed formula (\ref{i11}) and introduced the
notation $\nu^2=I^2_{10}/(I_{00}N(0))$.  Setting $\omega=i\Omega$,
Eq.~(\ref{deq}) becomes
\beq
0.5\ln{\epsilon_c\over \Omega}={1\over \lambda}-{\nu^2\over \kappa+B\Omega^2\ln(\epsilon_c/\Omega)},
\label{ok}
\eeq
with $B=(\partial^2 B(\Omega)/\partial \Omega^2)_0/\ln(\epsilon_c/\Omega)>0$. The sign of $B$ is
readily established upon replacing $\omega\to i\Omega$ in
Eq.~(\ref{prbcs}).

Let us now characterize the solutions of this equation in different quadrants
of the Lifshitz plane ($\lambda,\kappa$), while acknowledging that for any
realistic $nn$ interaction potential these parameters are constrained by
one another.  In the major part of the first quadrant ($\lambda>0,\kappa>0)$,
the magnitude of the term proportional to $\nu^2$ is suppressed due to
the poor overlap between the functions $\phi(p)$ and $\zeta_0(p)$
entering the integral $I_{10}$. Consequently, the role of this
term reduces to a renormalization of $\lambda$, leaving us with
the single BCS solution (\ref{bcsres}).

As already indicated, the function $\kappa(\rho)$ becomes negative at
$\rho<\rho_t$, {\it triggering the onset of the dineutron state}.
Furthermore, in the quadrant $(\lambda>0,\kappa<0)$, which is relevant
to the Argonne case at densities $\rho$ below about $0.4\rho_0$,
Eq.~(\ref{ok}) has {\it two different solutions}.  It is instructive
to trace the trajectories of both the roots $\Omega_{1,2}$ versus $\lambda$.
In the limit $\lambda\to 0$, the left root closest to the origin behaves as
  $\Omega_1\propto e^{-2/\lambda}$,
with $\Omega_1(0)=0$.  It should
therefore be identified with the BCS-like root.  In the limit addressed,
the other root
  $\Omega_2$ occurs close to $\sqrt{|\kappa|/B}$.
This root is
thus definitely of non-BCS nature.  As $\lambda$ increases, both roots
move away from the origin.  It must be stressed that in our analysis both
of the parameters $\lambda$ and $\kappa$ are supposed to be small to ensure
the smallness of the roots; in this respect the analysis is self-consistent
and implies that the inequality $\Omega_1(\lambda)<\Omega_2(\lambda) $
holds as the roots evolve.
This inequality implies that the non-BCS scenario ensures a shorter
relaxation time for the rearrangement of the normal
state than the standard BCS one does that obviates the latter scenario.

In the third quadrant $(\lambda<0,\kappa<0)$, the right non-BCS root
{\it no longer exists}, but the left one survives.  This root may be
treated as an unconventional BCS root in the following sense.  It
is true in a significant domain of the quadrant, the BCS-like
behavior \cite{kkc}
  $\Omega\propto e^{-2/\lambda_{\rm eff}}$ applies, with
  $1/\lambda_{\rm eff}=\nu^2/|\kappa|-1/|\lambda|$.
However, such behavior
is completely rearranged near the critical point $\lambda= 0$, where it
becomes {\it non-exponential}:
  $\Omega(\lambda\to 0) \to \sqrt{|\kappa|/B\ln(\epsilon_cB/|\kappa|)}$.

\section{Competition between BCS and in-medium dineutron correlations at
finite temperature}

In this section, we examine the temperature evolution of the two
different types of pairing correlations revealed in the quadrant
$(\lambda>0,\kappa<0)$.  We focus in each case on a possible
phase transition that occurs at the critical temperature
for termination of pairing correlations of the given type, as
determined from the Thouless criterion.\cite{thouless,trio}
This criterion takes the form of a linear integral equation
\beq
{\cal T}(p,T)=-\int V_0( p,p_1){\tanh\left(\epsilon(p_1)/2T\right)
\over 2\epsilon(p_1)}{\cal T}(p_1,T)d\upsilon_1
\label{thc}
\eeq
analogous to Eq.~(\ref{gc}) explored in Sec.~III.

\begin{figure}
\scalebox{0.50}
{\includegraphics{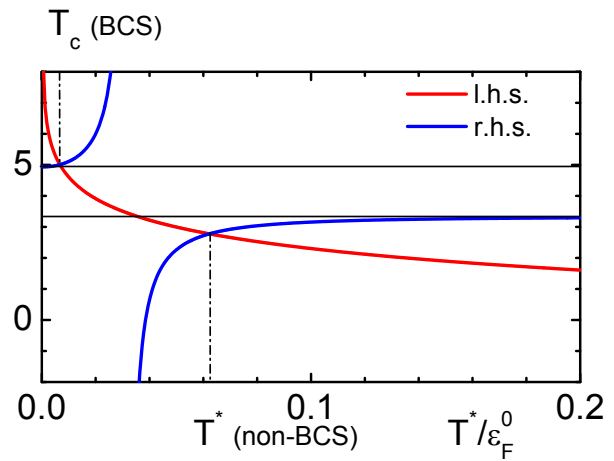}}
\caption{Graphical illustration of two solutions of Eq.~(\ref{dists}).
The left side (l.h.s.) of this equation as a function of $T^*$ measured
in units of $\epsilon_F^0=p_F^2/2M$ is shown by the dashed line, while the
right side (r.h.s.) is drawn as two solid curves. The following set of model
parameters is used: $\lambda=0.3$, $\nu=0.04$, $\kappa=-0.001$,
$\gamma=(\epsilon_F^0)^{-1/2}$, $\epsilon_c=\epsilon_F^0$.
The two crossing points yield two critical temperatures, $T_c$ and $T^*$
with $T^*>T_c$. Thus it is the in-medium dineutron solution that is
responsible for the maximum critical temperature for termination of pairing
correlations.}
\label{figure_disp_eq}
\end{figure}

To proceed further it is instructive to analyze this equation with
the aid of the decomposition procedure introduced in Sec.~III.
With details relegated to the Appendix, we proceed immediately
to the final result
\beq
0.5\ln (\epsilon_c/T)=1/\lambda-{\nu^2\over \kappa(\rho)+\gamma(T)}
\label{dists}
\eeq
where
\beq
\gamma(T)=-{1\over I_{00}}\int\zeta_0(p) \left({\tanh {\epsilon(p)\over 2T}
\over 2\epsilon(p)}-{1\over 2|\epsilon(p)|}\right)\zeta_0(p)d\upsilon.
\eeq
This expression can be recast as
\beq
\gamma(T)\propto \int\limits_0^{\infty}\epsilon
{e^{-\epsilon/T}d\epsilon\over e^{\epsilon/T} +1}=\gamma T^2 .
\eeq
The graphical solution of Eq.~(\ref{dists}) is shown in
Fig.~\ref{figure_disp_eq}, the two sides of this equation being plotted
versus the temperature.
  The logarithmic curve is seen to cross
both branches of the hyperbolic curve, and hence Eq.~(\ref{dists}) does
possess two roots $T_c$ and $T^*$.  In the limit $\lambda\to 0$, we
naturally obtain the BCS solution $T_c\propto  e^{-1/\lambda}$.  Even so,
in the region $\lambda>0,\kappa<0$ there exists another,
non-BCS root of Eq.~(\ref{dists}) written as
\beq
0.5\ln (\epsilon_c/T)=1/\lambda+{\nu^2\over |\kappa(\rho)|-\gamma T^2},
\label{dit}
\eeq
namely $T^*(\lambda\to 0)=\sqrt{|\kappa(\rho)|/\gamma}>T_c$
 that has no the exponential smallness.
   This result
informs us that in the region $\lambda>0,\kappa<0$, the BCS solution
{\it loses the competition with the in-medium dineutron solution}
not only in the interval $T_c<T<T^*$ where the BCS solution does not exist
 but also at $T<T_c$ where BCS gain in energy, being exponentially small, ranks below the dineutron one
  in the value.

In the quadrant $\lambda<0,\kappa<0$, the right non-BCS root is seen
to disappear, and there remains a single bizarre solution which, at
  $|\kappa|<\nu^2|\lambda|$,
behaves in harmony with the BCS-like formula
  $T_c\propto e^{-2/\lambda_{\rm eff}}$,
in which
  $1/\lambda_{\rm eff}=\nu^2/|\kappa|-1/|\lambda|$.
Otherwise, however, this solution exhibits
non-BCS behavior with $\Omega(T^*)\simeq \sqrt{|\kappa|/\gamma}$.

\section{Phase transition between BCS and pseudogap states in strongly
correlated Fermi systems at finite temperature}

In principle, the above results, derived employing the standard Fermi-liquid
spectrum $\epsilon(p)\propto (p-p_F)$, need not hold in general, and most
especially when the system is subject to very strong correlations in the
particle-hole channel.  Such correlations often give rise to a so-called
quantum critical point (QCP) where the effective mass $M^*$ diverges.
This behavior triggers a rearrangement of the Landau state\cite{zb,shagp}
which, in its turn, alters the left side of the dispersion equation,
thereby affecting its roots and their relationship.

An especially profound change occurs in the event of the collapse of
the Fermi surface at a finite temperature $T_M$, beyond which
no BCS solution can possibly exist.\cite{khodel}  This situation
warrants more detailed discussion.  In conventional homogeneous
Fermi liquids, the equation
\beq
\epsilon(p)=0,
\label{top}
\eeq
which is to be satisfied by the $T=0$ single-particle spectrum $\epsilon(p)$,
has the single root $p=p_F$, which locates the Fermi surface.  The same is
true at finite $T$ as well, up to some critical temperature $T_M$ where
this root disappears. The function $\epsilon(p)$ then becomes positive
definite, implying that the Fermi surface collapses at this temperature,
marking the onset of classical physics.\cite{khodel}  Correspondingly,
the left side of Eq.~(\ref{ok}) ceases to be logarithmically divergent,
and the customary BCS solution, existing at small $\lambda$, disappears.
In conventional Fermi liquids having effective masses $M^*$ not so
different from the bare mass $M$, the collapse of the Fermi surface
occurring at $T_M\simeq\epsilon^0_F$ has no impact on BCS correlations,
since BCS superconductivity has already been terminated well
before the collapse takes place.

\begin{figure}
\scalebox{0.48}
{\includegraphics{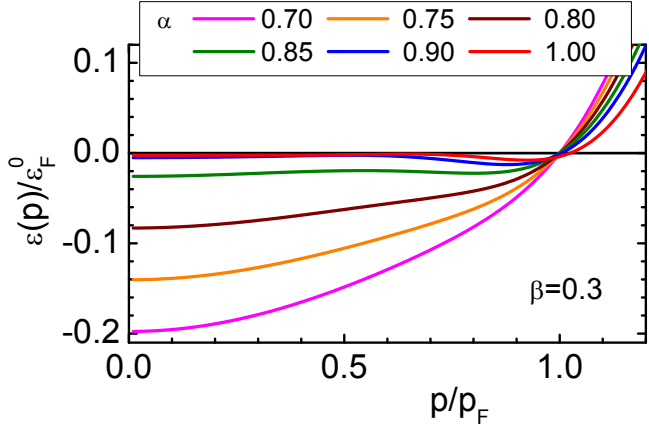}}
\caption{Single-particle spectra in the model of homogeneous fermion
matter with quasiparticle interaction (\ref{model}), plotted
for $\beta=0.3$ and different values of $\alpha$.}
\label{figure_kappa_03}
\end{figure}

However, in strongly correlated Fermi systems lying on the edge of
stability of the Landau state, the form of the spectrum $\epsilon(p)$
can be completely different than in FL theory.  This is illustrated
in Fig.~\ref{figure_kappa_03}, which presents results of numerical
calculations of the single-particle spectrum $\epsilon(p)$, as
determined within a model described in Ref.~\onlinecite{prb2008}.
The Landau interaction function $f(q)$ of this model has the
dimensionless form
\beq
f(q)N(0)= {\alpha\over q^2+\beta^2p_F^2}
\label{model}
\eeq
with $\beta=0.3$ and different choices for $\alpha$.  The bandwidth
$W=|\epsilon(p=0)|$ is seen to shrink dramatically as the interaction
strength approaches a critical value, at which the stability of the
conventional Landau state with $n(p)=\theta(p_F-p)$ is lost and the
Fermi surface becomes multi-connected.\cite{zb,shagp}   In the
vicinity of the critical point, the single root $p_F(T)$ of
Eq.~(\ref{top}) at which the spectrum $\epsilon(p,T)$ changes sign
is found to approach zero rapidly as the temperature $T$ is raised.
This behavior contrasts sharply with that seen in the conventional
FL case, where this root moves extremely slowly: $p_F(T)=
p_F(0)+O\left(T^2/(\epsilon^0_F)^2\right)$.

\begin{figure}
\scalebox{0.6}
{\includegraphics{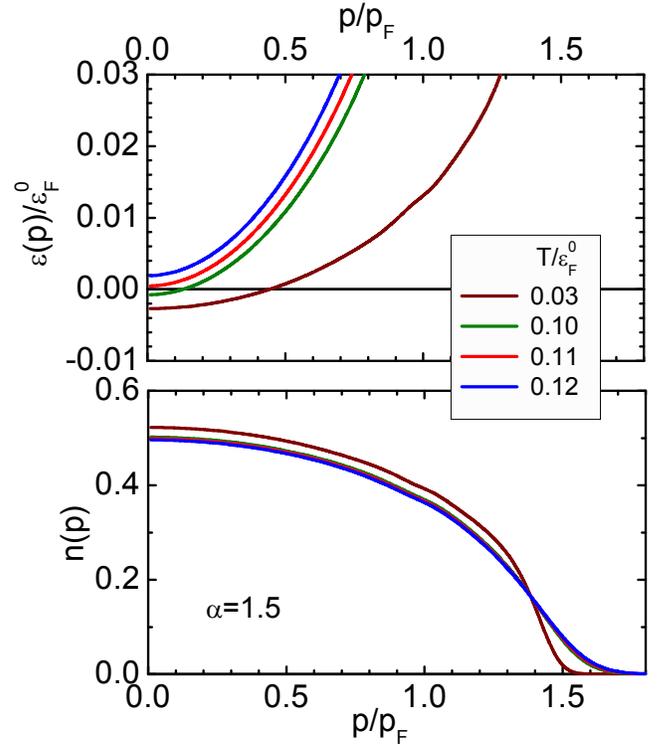}}
\caption{Behavior of the single-particle spectrum (shown in top panel in units
of $\epsilon_F^0$) and quasiparticle momentum distribution (bottom panel)
in a super-strongly correlated Fermi system in which all the quasiparticles
reside in the fermion condensate. The model (\ref{model}) is assumed
with $\beta=0.07$, $g=1.5$.
}
\label{figure_3dsf_35}
\end{figure}

\begin{figure}
\scalebox{0.6}
{\includegraphics{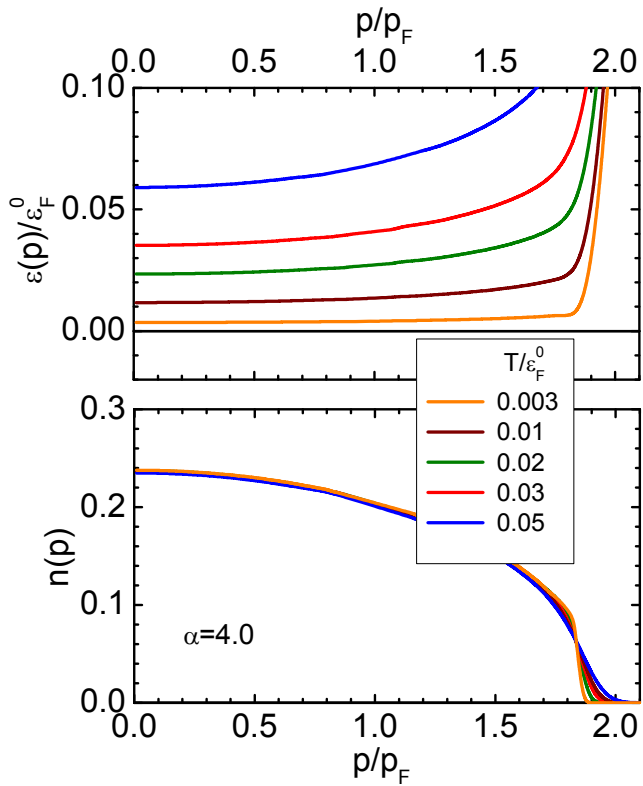}}
\caption{The same as in Fig.~\ref{figure_3dsf_35} but for $g=4$.}
\label{figure_3dsf_100}
\end{figure}

Suppose now that there exist in nature homogeneous Fermi systems that
are so strongly correlated that all quasiparticles go into the fermion
condensate (FC), consisting of the totality of single-particle states
belonging to a completely flat spectrum.  (See
Refs.~\onlinecite{shagrep,mig100,an2012} for comprehensive
reviews of this phenomenon and its implications, as well as
Ref.~\onlinecite{prb2008}.)  Illustrative results from numerical calculations
are presented in Figs.~\ref{figure_3dsf_35} and Fig.~\ref{figure_3dsf_100},
which display single-particle spectra $\epsilon(p)$ evaluated for
different temperatures based on the same form (\ref{model}) for the
quasiparticle interaction, but with different input parameters
($\beta = 0.07$ and $\alpha = 1.5$ and $4.2$, respectively).  The
coupling constant $\alpha=1.5$ is so large that all the quasiparticles
reside in the fermion condensate.  At zero temperature, we find
$\epsilon(p=0) \simeq  -0.003\epsilon_F^0$, whereas at
$T_M\simeq 0.1\epsilon_F^0$, this quantity changes sign,
so that at $T>T_M$ roots of Eq.~(\ref{top}) no longer exist.
Fig.~\ref{figure_3dsf_100} demonstrates the behavior of the spectra
within the same model (\ref{model}), but at $\alpha=4.2$.  These
results inform us that at $\alpha=1.5$ the spectrum $\epsilon(p)$
assumes negative values only at $T<0.05\epsilon^0_F$, and only over
a small range at small $p$, while at $\alpha=4.2$ the function
$\epsilon(p)$ becomes positive independently of both momentum $p$
and temperature $T$.  Thus, such an super-strongly correlated
Fermi system {\it has no Fermi surface at all}.  Evidently, if in
this case the Thouless equation (\ref{thc}) has a nontrivial
solution, it must be attributed to the pseudogap phase.

An examination of the spectra $\epsilon(p)$ drawn in
Figs.~\ref{figure_3dsf_35} and \ref{figure_3dsf_100}
allows one to infer that beyond $T_M$, the behavior of the integral
  $I_{11}(T)$
evolves from the conventional BCS logarithmic character, in
which both solutions survive, to a behavior almost independent of $T$.

In the latter case, we are left with a single solution, which
is the non-BCS solution provided $L<1/\lambda$ or
a BCS solution provided $L>(1/\lambda+\nu^2/|\kappa|)$;
otherwise, nontrivial solutions of the dispersion equation
(\ref{ok}) do not exist at all. In this statement, the notation
$L$ is employed for a posited value of the integral $I_{11}$ in
the case of non-Fermi-liquid behavior of the single-particle
spectrum $\epsilon(p)$.  Referring to Fig.~\ref{figure_disp_eq},
the three situations just identified correspond to $L$ lying
(i) below the lower horizontal line (non-BCS),
(ii) above the upper horizontal line (BCS),
and (iii)
between the two horizontal lines (no nontrivial solution).

\section{Discussion and Conclusions}

For over fifty years, theoretical consideration of pairing correlations
in neutron matter has been carried out uniquely within BCS theory.
No work based on some other conception of pairing correlations
has appeared hitherto, despite the salient fact that in a significant
density domain, the pairing constant $\lambda$ has the ``wrong''
(i.e., negative) sign with respect to BCS theory. This article is the
first to treat on equal footing the BCS pairing correlations
giving rise to the Cooper condensate and the non-BCS pairing
correlations that induce in-medium dineutron formation.

Let us summarize the findings of our present analysis of the pairing
instabilities of the normal state of neutron matter, which may serve as
a guide to resolution of the remaining qualitative and quantitative issues
that have been been exposed.  The arguments offered and results obtained
demonstrate that contrary to common belief, the Lifshitz phase diagrams of
neutron matter and other comparable many-fermion systems that are subject
to pairing correlations are characterized by {\it two} dimensionless
parameters, associated with two different pairing scenarios that operate
in different density regions.  The first of these, $\lambda=-V(p_F,p_F)N(0)$,
represents the diagonal matrix element of the pairing interaction evaluated
on the Fermi surface.  This parameter, associated with Cooper pairing of
quasiparticles whose energies lie very near the Fermi surface, is relevant
in the case $\lambda>0$.

The second parameter characterizing the
Lifshitz phase diagram, denoted by $\kappa$, has scant relation
to the Fermi surface.  The sign of this parameter is indicative of the
possibility of dineutron formation in the medium.  If $\kappa$ is positive,
the dineutron correlations interfere with the BCS mechanism so as to suppress
conventional pairing correlations. If $\kappa$ is negative, as expected to
apply in neutron matter over some density range below $\rho_t\simeq 2\rho_0$,
the system is able to undergo a dineutron phase transition analogous to
the formation of bipolarons in solid state physics.  Correspondingly,
as we have seen, two different phases organized by pairing correlations
a conventional BCS phase with a critical temperature $T_c$ for termination of the
Cooper condensate, and an in-medium dineutron phase.

In the customary situation where the neutron spectrum has the Fermi-liquid
form $\epsilon(p)=p_F(p-p_F)/M^*$ with $M^*\simeq M$, the critical temperature
$T^*$ for dissolution of dineutron correlations exceeds the temperature
$T_c$ beyond which the BCS scenario becomes scenario irrelevant.  Indeed,
near $T_c$ the BCS gain in energy is extremely small, being proportional
to $(T_c-T)^2$, while the corresponding shift in energy due to dineutron
correlations contains no such small factor.  As for the quadrant
$(\lambda<0,\kappa<0)$, only a hybrid phase survives, which, at $|\kappa|<\nu^2|\lambda|$, exhibits a BCS-like behavior with $\lambda_{\rm eff}>0$, while it becomes a non-BCS phase in the limit $\lambda\to 0$. Even
such a restricted analysis as we have performed demonstrates that there
is little or no room for conventional BCS correlations in neutron matter.

In some respects, the situation created by these revelations of the
nature of pairing correlations in neutron matter is reminiscent of
that which arose for the low-lying $2^+$-collective oscillations of
atomic nuclei at the dawn of the age of the standard nuclear paradigm.
At that time, commonly employed models with simple effective
nucleon-nucleon interactions of the quadrupole-quadrupole type
entailed a description of these excitations as quanta of zero sound
of the bulk nucleus, a volume effect.  While agreement of the
theoretical results with the experimental data on collective frequencies
was achieved, the posited nature of the $2^+$ levels turned out to be
incorrect.  Indeed, later developments,
corroborated by experimental
data on the transition densities, measured in inelastic scattering of high-energy electrons, have established
that these oscillations belong instead to the Goldstone surface mode
associated with loss of translational invariance \cite{khodjetpl,bertsch,khodyaf,physrep,plat}.
  In the
nuclear pairing problem, model calculations with effective forces having
positive pairing constant $\lambda_{\rm eff}$ have gained widespread
acceptance as well, implying the presence of a Cooper condensate.
 In our article we argue that this scenario has flaws.
Alas, as yet there is no a distinct experimental method for the
measurement of the transition density in the Cooper channel. Furthermore,
even there were, significant difficulties would be encountered in the
interpretation of corresponding experimental data because, unlike the
case of low-lying collective excitations of atomic nuclei, there is
no blatant contradiction between the BCS and non-BCS transition densities.

Therefore we adopt a different strategy,
based on comparison of the energy shifts $\delta E_0$ associated with the
onset of the pairing correlations in the BCS and  in-medium
dineutron scenarios where the corresponding phase transition belongs to a family of
second--order phase transitions, whose properties are properly explained  within the
theory of second--order phase transitions.
   In light of this situation, the analogy with the problem
of the spontaneous quadrupole deformations of atomic nuclei
is helpful.  The theory of nuclear deformations provides
the formula
\beq
\delta E_0=C\beta^2+ D\beta^4+ \ldots  ,
\label{gain}
\eeq
containing the deformation parameter $\beta$, the stiffness coefficient $C$, and the coefficient $D$ (presumed positive), which is responsible for repulsive interactions between the collective quadrupole modes. On the disordered side of the phase transition, the stiffness $C$ has a positive value, so that $\beta=0$.  Beyond the phase transition point, $C$ changes sign, triggering the emergence of a new phase with the deformation parameter $\beta^2_0=-C/2D$ and the shift in ground state energy $\delta E_0=-C^2/4D$.
Definitely, near the  new equilibrium point where, as seen, $\delta E_0(\beta)\propto C_{\beta}(\beta-\beta_0)^2$, the stiffness coefficient  $C_{\beta}$ turns out to be positive. This is similar to what happens for the first term of the expansion of the thermodynamic potential in the theory of second-order phase transitions.

It is quite significant that formulas analogous to Eq.~(\ref{gain}) appear
not only in the theory of nuclear deformation but also in the Landau
theory of second-order phase transitions, in the theory of pion
condensation,\cite{migdalp}
in the self-consistent theory of low-energy nuclear phenomena,\cite{physrep}
and in many refined versions of mean-field theory (see
e.g.~\onlinecite{sigrist,kee}).  In all these theories, the stiffness
coefficient $C$ is expressed unambiguously in terms of the inverse
response function $\chi^{-1}$ by means of the formula relating $\chi$
to the variation of the ground state energy.  This relation, written
symbolically as $\delta E_0=(1/2)\chi \delta v_0\delta v_0\equiv
(1/2)\chi^{-1} \delta \rho\delta \rho$, is to be evaluated on the
disordered side of the transition and then applied on the ordered
side.  (By definition, $\chi=\delta\rho/\delta v_0$, where
$\delta \rho$ is the density variation produced by a weak static
external field $\delta v_0$.)

In the nuclear pairing problem, the corresponding response function
$\chi_C$ should be evaluated in the Cooper channel,
implying that  $(\chi_C)_{\alpha\beta}=\chi_C(\tau_2)_{\alpha\beta}$;
This can be done with the aid of the same decomposition strategy (\ref{decg})
deployed earlier in  this text.  Omitting intervening mathematical
steps, we give the final result
\beq
\chi_C(p)= L(p,0) {\cal T}(p)\propto \kappa^{-1} L(p,0)\zeta_0(p)   .
\eeq
Accordingly, the inverse response function $\chi_C^{-1}$, and hence
the stiffness coefficient entering as the first term of the
expansion of the energy shift $\delta E_0$ of the pairing problem,
turn out to be proportional to the critical quantity $\kappa(\rho)$,
which becomes negative on the ordered side of the dineutron
phase transition.    Other than assuming it to
be positive, we do not consider here  the second term $D$ of
the corresponding expansion, which is proportional to the dineutron
scattering amplitude.  In this case, the energy shift $\delta E_0$
increases linearly with $\kappa^2(\rho)$ as one moves farther
from the point of the dineutron phase transition.  If the assumption
$D>0$ fails, the expansion should, as usual, be supplemented by
successive terms.  At any rate, the shift $\delta E_0$ {\it is not}
exponentially small.

Contrariwise, within the established framework of BCS theory, the
BCS energy shift near the critical density $\rho_t$ is of course
exponentially small\cite{kkc}:
\beq
\delta E_0^{BCS}(T=0)\propto \Delta^2_0\propto e^{-a\rho_t/\kappa(\rho)},
\eeq
with $a$ as a numerical factor.  Conclusively, the dineutron effect
wins the energetic competition for ascendancy.

Our analysis has been restricted to the vicinity of the critical points,
where presumably the second-order phase transition scenario we have
employed is applicable. In obtaining all our results, we have proceeded from the assumption, commonly adopted in work on the nuclear pairing problem, that the pairing
interaction is given by the in-vacuum or ``bare'' interaction potential $V$.
This assumption is clearly of limited validity, because medium effects
can significantly alter the pairing interaction from its vacuum counterpart.
Unfortunately, the associated ``polarization corrections'' to the
pairing interaction still await proper investigation.  In this connection
it is worth remembering that Pankratov et al.\cite{edik2} have shown
that the gap values in atomic nuclei are overestimated by a factor
two when such renormalization effects are ignored.  One of explanations of this discrepancy is associated with a complete suppression of pairing correlations in the nuclear interior that renders the pairing correlations the surface phenomenon.\cite{baldo}
This situation highlights the timeliness of
the message of S.~T.\ quoted in the Preamble, confirming the wisdom spoken by a
great Russian poet
Fyodor Tyutchev more than 100 years ago: ``Ќ ¬ ­Ґ ¤ ­® ЇаҐ¤гЈ ¤ вм, Є Є б«®ў® ­ иҐ ®в§®ўҐвбп \dots''\footnote{``We cannot know further ways of our word --- how it'll be drifted \dots''}

\section*{Acknowledgments}

We thank M.~Alford, S.~Pankratov and E.~Saperstein for fruitful discussions.
This work was partially supported by the Grants NSh-215.2012.2
and 2.1.1/4540 of the Russian Ministry for Science and Education
and by the RFBR Grants 11-02-00467 and 13-02-00085, as well
as the McDonnell Center for the Space Sciences. JWC expresses
his gratitude to Professor Jos\'e Lu\'is da Silva and his
colleagues at Centro de Ci\^encias Matem\'aticas for kind
hospitality and extensive discussions during summer residence
at the University of Madeira.

\section*{Appendix}

In this appendix we recast the Thouless equation (\ref{thc}) into a form
similar to that obtained in the analysis of the instabilities of
the normal state.  To begin, the pairing interaction $V_0$
is decomposed as
\beq
V_0(p_1,p_2)= V_F\phi(p_1)\phi(p_2)+ {\cal R}(p_1,p_2) ,
\label{dec}
\eeq
where $\phi(p)= V_0(p,p_F)/ V_F$, with $ V_F= V_0(p_F,p_F)$.
Upon inserting Eq.~(\ref{dec}) into Eq.~(\ref{thc}) and performing
some algebra, Eq.~(\ref{thc}) is transformed into a set of two
equivalent equations, the first of which is given by
\beq
\chi(p)=\phi(p)-\int {\cal R}( p,p_1) {
\tanh\left(\epsilon(p_1)/2T^*\right)
\over 2\epsilon(p_1)}\chi(p_1)d\upsilon_1,
\label{chit}
\eeq
while the second takes the form
\beq
-{1\over  V_F}=\int \phi(p){
\tanh\left(\epsilon(p)/2T^*\right)
\over 2\epsilon(p)}\chi(p)d\upsilon.
\label{tempt}
\eeq
After introducing the difference $\eta(p)=\chi(p)-\phi(p)$ and
taking several algebraic steps one arrives at
\beq
-{1\over V_F}= I_{11}(T^*) +\int \phi(p)
{\tanh\left(\epsilon(p)/2T^*\right)
\over 2\epsilon(p)}\eta(p)d\upsilon  ,
\label{temrt}
\eeq
where
\beq
I_{11}(T)=\int \phi(p){\tanh {\epsilon(p)\over 2T}
\over 2\epsilon(p)}\phi(p)d\upsilon\simeq
  0.5N(0)\ln (\epsilon_c/T),
\label{intt}
\eeq
while the function $\eta(p)$ obeys equation
\beq
\eta(p,T)=
-\int {\cal R}( p,p_1) {\tanh {\epsilon(p_1)\over 2T}
\over 2\epsilon(p_1)}\left(\phi(p_1)+\eta(p_1,T)\right)d\upsilon_1 .
\label{et}
\eeq
We observe that at the bifurcation point $T^*=0$, the first term on the
right side of Eq.~(\ref{temrt}) diverges logarithmically, so that a solution
$T^*=0$ exists only if the second term also {\it diverges} at this point.
To confirm that the latter is the case, we expand the function $\eta(p)$
in a basis formed by the eigenfunctions $\zeta_n(p)$.  Extracting
the main term proportional to $\zeta_0(p)$ explicitly, we write
\beq
\eta(p,T)=\eta_0(T)\zeta_0(p)+\vartheta(p),
\label{exot}
\eeq
where, as before, the eigenfunction $\zeta_0(p)$ obeys the equation
\beq
\zeta_0(p)=-\sigma_0\int {\cal R}( p,p_1,\rho){1\over 2|\epsilon(p_1)|}
\zeta_0(p_1) d\upsilon_1 ,
\label{ztoa}
\eeq
while the remainder $\vartheta(p)$ vanishes at the Fermi surface like
$\zeta_0(p)$ and $\eta(p)$.  Upon inserting this expansion into
Eq.~(\ref{exot}) and collecting all terms containing the factor
$\eta_0(T)$ on the left side of Eq.~(\ref{et}), one obtains
\beq
\eta_0(T)\left (\zeta_0(p)+ \int {\cal R}( p,p_1)
{\tanh {\epsilon(p_1)\over 2T}
\over 2\epsilon(p_1)}\zeta_0(p_1)d\upsilon_1\right)=Z(p) ,
\label{etaz}
\eeq
where
\beq
Z(p)=-\vartheta(p)-\int {\cal R}( p,p_1) {\tanh {\epsilon(p_1)\over 2T}
\over 2\epsilon(p_1)}\left(\phi(p_1)+\theta(p_1)\right)d\upsilon_1 .
\eeq
The left side of Eq.~(\ref{etaz}) is conveniently rewritten with the aid
of Eq.~(\ref{ztoa}) to yield
\beq
\eta_0(T)\left ({\kappa\over \sigma_0}\zeta_0(p)+
\int {\cal R}( p,p_1){\cal D}(p_1,T) \zeta_0(p_1)d\upsilon_1\right)=Z(p) ,
\label{eqo}
\eeq
where
\beq
{\cal D}(p,T)={
\tanh\left(\epsilon(p)/2T\right)
\over 2\epsilon(p)}-{1\over 2|\epsilon(p)|} .
\eeq
Next, both sides of this equation are multiplied by the product
$\zeta_0(p)/(2|\epsilon(p)|)$, the momentum integration is performed.
Eliminating the operator ${\cal R}$ with the aid of Eq.~(\ref{ztoa}),
we arrive at
\beq
\eta_0(T)=(\kappa+\gamma(T))^{-1}I_{10}/I_{00},
\label{rt}
\eeq
where $I_{00}>0$ is given by Eq.~(\ref{i00}) of Sec.~III, and
\beq
  \gamma(T)=-I_{00}^{-1}
\int\zeta_0(p) \left({\tanh {\epsilon(p)\over 2T}
\over 2\epsilon(p)}-{1\over 2|\epsilon(p)|}\right)\zeta_0(p)d\upsilon <0 ,
\label{gammt}
\eeq
while
\beq
I_{10}=\sigma_0\int \zeta_0(p){1\over 2|\epsilon(p)|} Z(p) d\upsilon .
\label{i10}
\eeq
To demonstrate that the sign of $\gamma(T)$ is indeed positive, we rewrite
the integrand of Eq.~(\ref{gammt}) according to
\beq
\gamma(T)\propto \int\zeta_0^2(p){1-\tanh ({|\epsilon(p)|/ 2T})
\over 2|\epsilon(p)|} d\upsilon .
\eeq
Since $\zeta_0(p)$ vanishes at the Fermi surface like $\epsilon(p)$, we
immediately conclude that $\gamma(T)=\gamma T^2$.
Considering now the integral $I_{10}$, the explicit form of $Z(p)$ is
inserted into the integrand of Eq.~(\ref{i10}, and it is verified that
the terms involving the remainder $\vartheta$ practically cancel each
other.  Accordingly, we may take
\beq
I_{10}=\int \zeta_0(p) {1\over 2|\epsilon(p)|}\phi(p)d\upsilon .
\eeq
Since $\gamma(T)$ vanishes at $T\to 0$, the coefficient $\eta_0(0)$ does
in fact diverge together with the function $\eta(p)$ itself\cite{kkc}
at the critical point where $\kappa$ changes sign.

Next we substitute Eq.~(\ref{rt}) into the dispersion equation (\ref{temrt})
to find
\beq
0.5\ln (\epsilon_c/T)=1/\lambda-{\nu^2\over
  \kappa(\rho)+\gamma T^2},
\label{okt}
\eeq
with the understanding that all minor corrections are included in the effective
pairing constant denoted $\lambda$, as before.  Near the critical density
$\rho_t$ the parameter $\kappa$, being negative, behaves as
$-|\left(\partial \kappa/\partial \rho\right)_t|(\rho_t-\rho)$.
The term $1/\lambda$ in Eq.~(\ref{okt}) can then be omitted, and we
are left with\cite{kkc,khod1997}
\beq
T_c(\rho)\propto e^{-a\rho_t/(\rho_t-\rho)},
\label{tccra}
\eeq
where $a$ is a numerical constant.  On the other hand, Eq.~(\ref{okt}) has
the the second, non-exponential root
\beq
T^*\simeq \sqrt{|\kappa|/\gamma}
\eeq
corresponding to the dineutron phase.  We conclude that $T^*>T_c$, and
consequently that the BCS solution loses the competition with the
dineutron solution.

\end{document}